# CONVERSATIONAL RECOMMENDATION SYSTEM USING NLP AND SENTIMENT ANALYSIS

[1]PIYUSH TALEGAONKAR, [2]SIDDHANT HOLE, [3]SHRINESH KAMBLE, [4]PRASHIL GULECHHA, [5]DEEPALI SALAPURKAR

[1,2,3,4]Student, Pune Institute of Computer Technology, Dhankavadi, Pune
[5]Assistant Professor, Pune Institute of Computer Technology, Dhankavadi, Pune
E-mail: [1]talegaonkar.piyush@gmail.com, [2]holesiddhant@gmail.com, [3]shrineshkamble2002@gmail.com,
[4]prashilgulechha@gmail.com, [5]dpsalapurkar@pict.edu

**Abstract** - In today's digitally-driven world, the demand for personalized and context-aware recommendations has never been greater. Traditional recommender systems have made significant strides in this direction, but they often lack the ability to tap into the richness of conversational data. "Conversational Recommender System for Marketing Application using Deep Learning," represents a novel approach to recommendation systems by integrating conversational insights into the recommendation process. The Conversational Recommender System integrates cutting-edge technologies such as deep learning, leveraging machine learning algorithms like Apriori for Association Rule Mining [1], Convolutional Neural Networks (CNN), Recurrent Neural Networks (RNN), and Long Short-Term Memory (LTSM) [10]. Furthermore, sophisticated voice recognition technologies, including Hidden Markov Models (HMMs) and Dynamic Time Warping (DTW) algorithms [13], play a crucial role in accurate speech-to-text conversion, ensuring robust performance in diverse environments. The methodology incorporates a fusion of content-based and collaborative recommendation approaches, enhancing them with NLP techniques. This innovative integration ensures a more personalized and context-aware recommendation experience, particularly in marketing applications.

## I. INTRODUCTION

### 1.1 Background

Recommendation systems are everywhere – from entertainment to healthcare, education, E-Commerce and so on. With the exponential rise of big data, new and advanced recommendation systems are bound to be developed. Technology has evolved to such a stage that merely the activity of the user and the similarity between the items may not be enough for appropriate and efficient recommendation. In the modern era of technology, people's conversations and subconscious behaviors would also factor in during the process of recommendation by an artificially intelligent system.

### 1.2 Motivation

Traditionally, the recommendations are made based on the activity of the user, computed using Machine Learning [2] or Deep Learning models [6]. It involves elements of text processing and analysis. Therefore, recommendations can also be done through voice capturing of the user, thereby increasing the marketing purpose, and putting a step forward in the recommendation systems currently present.

## II. LITERATURE SURVEY

1) **Machine Learning Based Recommender System for E-Commerce [1]:**
Authors: Manal Loukili, Faycal Messaoudi, Mohammad El Ghazi.
Year of Publication: 2023.
Summary: A practical implementation of a Machine Learning based Recommendation system using Association Rule Mining.

2) **Big Data cloud-based recommendation system using NLP techniques with machine and deep learning [3]**
Authors: Hoger K. Omar, Mondher Frikha, Alaa Khalil Jumaa
Year of Publication: 2023
Summary: This research paper presents a cloud-based recommendation system for processing big data, employing matrix factorization with three distinct approaches, including traditional Singular Value Decomposition (SVD), Apache Spark's Alternating Least Squares (ALS) algorithm, and deep learning with TensorFlow.

3) **Recommender Systems in E-Commerce [5]**
Authors: Lopamudra Mohanty, Laxmi Saraswat, Puneet Garg, Sonia Lamba.
Year of Publication: 2022
Summary: A brief explanation of different types of Recommender Systems used in E-Commerce applications, including Collaborative, Content-Based, Knowledge-Based and Community-Based, among other subtypes.

4) **A Deep Neural Network (DNN) Approach for Recommendation Systems [10]**
Authors: Shashi Shekhar, Anshy Singh & Avadhesh Kumar Gupta
Year of Publication: 2022

Summary: This research paper introduces a recommendation system model that combines deep learning technology with Collaborative Filtering (CF) algorithms to improve recommendation quality in the face of data sparsity and scalability challenges. By effectively addressing issues related to user behaviour and feature analysis, the proposed model demonstrates superior performance compared to traditional techniques, as evidenced by higher metrics such as F1-measure, Recall, and Precision in experimental evaluations using Amazon dataset.

**5) A survey on Conversational Recommendation systems [11]**
Authors: Dietmar Jannach, Ahtsham Manzoor, Wanling Cai, Li Chen.

Year of Publication: 2021
Summary: A recommender system takes the dataset from any user search and recommends appropriate products. The dataset, when contains normal interactions and conversations, forms a conversational recommender system.

**6) Speech Recognition using Machine Learning [13]**
Authors: Vineet Vashisht, Aditya Pandey, Satya Yadhav.

Year of Publication: 2021
Summary: This research paper focuses on leveraging speech recognition technology to break down language barriers and enhance human-computer interactions. By implementing neural machine translation and speech synthesis, the study aims to enable efficient communication and control of devices through voice input, offering a potential solution for individuals facing language constraints.

## III. EXISTING METHODOLOGY

If one were to broadly classify existing recommendation system into two types [5], they would be:
- Content-Based [9]: In these systems, recommendation is made upon the user activity on the specific platform. The similarity is calculated by the association with the products themselves.
- Collaborative [7]: In these systems, recommendation is made based on the activities and ratings of the users like the main user.

Various machine learning [2] and deep learning [6] algorithms can be used to execute the recommendation process, including Apriori (Association Rule Mining), Convolutional Neural Networks (CNN), Recurrent Neural Networks (RNN), Long Short-Term Memory (LTSM) among other techniques [5].

The major purpose is to bridge the gap between traditional recommendation systems and the evolving landscape of conversational data. Leveraging advancements in deep learning, our methodology introduces a 'Conversational Recommender System' that seamlessly integrates user interactions and voice data into the recommendation process. The proposed system builds upon the foundation of content-based and collaborative recommendation approaches, augmenting them with the rich conversational insights obtained from user interactions within the application.

The existing methods of voice recognition systems incorporate a range of sophisticated technologies to enable accurate and robust speech-to-text conversion. These technologies include the utilization of Hidden Markov Models (HMMs) for analysing sound wave patterns [13], enabling the recognition of phonemes and words. Additionally, these systems leverage Dynamic Time Warping (DTW) algorithms, facilitating effective pattern matching between varying speed and time signals [15]. To address the challenge of noisy environments, voice recognition systems integrate Voice Activity Detectors, which employ techniques such as pitch detection, energy threshold analysis, periodicity measurement, and spectrum analysis to enhance the extraction of feature vectors during the speech recognition process.

Furthermore, advancements in computational capabilities have led to the incorporation of sophisticated neural networks that mimic the functioning of the human brain, enabling these systems to learn and recognize intricate speech patterns and nuances [6]. The comprehensive framework of these voice recognition systems encompasses a fusion of these technologies, enabling real-time and distributed speech recognition while addressing challenges such as diverse accents, varying speech speeds, and accurate punctuation recognition.

Existing research in the field of recommender systems points to the various types of algorithms and methods currently used. However, in the context of upcoming technologies, new types of recommender systems are prone to be deployed. One of those techniques, sparsely explored, is one of the conversational recommender systems [11]. It essentially means that recommendations are generated through the medium of real-life conversations. To put simply, traditional content-based [12] recommendation systems relied on the user activity on the application to generate the dataset for association functionality; however, in the new conversational recommender systems, even the regular conversations made by the user in day-to-day life will have tremendous impact in the quality of recommendations made by the application.

## IV. PROPOSED METHODOLOGY

As stated earlier, the recommendation model will be a prototype for an E-Commerce [5] application containing 10 product categories, and the prototype recommends the three most relevant products from the conversational input. The working of the model can be divided into three parts. The first one being voice recognition [13], the second one being the tokenization and filtering of important words, and the last one being the actual recommendation algorithm.

The stream of words, recognized after the user inputs it through the microphone of the client device, is passed forward for text analysis. The stream of words is separated into a collection of individual words. This process is called tokenization. A conversation may include a lot of unnecessary and filler words or phrases, which do not prove to be useful in the recommendation algorithm. So, to reduce the computational requirements and to increase the efficiency of the model, words that are not required are filtered out of the collection. This is done in two steps:
1) Stop Words: These are the filler words which do not add any semantic meaning to the sentence, like "a", "the" etc. These stop words are removed first from the list.
2) POS Tags: Parts-Of-Speech (POS) is a collection of the various types of words used in a sentence, in the "nltk" library. A few examples of the types of POS Tags are "Noun Singular," "Noun Plural," "List" etc. For the processing of the recommendation algorithm, only a few types of POS Tags are useful. Words having Tags other than those are filtered out of the list.

The filtered list of words is then passed on to the final stage of the model: the recommendation algorithm. The primary mechanism of the algorithm lies in word vectors [3] and similarities. The data is stored in a manner where each product category is a key of the object, and the value of the object consists of another object having its key as the keyword and the frequency of that keyword as the value. The algorithm compares the word vectors of the filtered list, first with the headings, and then with the keywords of the headings. Comparison is made using the cosine similarity function, which is helpful in determining the closeness of the spoken word with the relevant product names and keywords. The top 3 product categories having the highest weighted average of similarities are saved, and rechecked for each word in the list of the filtered input words. However, the top 3 categories extracted depend on the motivations and the sentiment of the speaker, which is calculated using sentiment analysis. The frequency of the keywords is taken into equation of the weighted average. This is how, the top 3 product categories are extracted after the processing of the model prototype.

## V. IMPLEMENTATION

The coding of the model prototype is done in python, and the custom E-Commerce application and website, used to demonstrate the working of the model prototype, is developed in flutter and React JS respectively. The product data is stored in a JSON object format, and is stored on the client side for personalization and additional security.

The implementation of the three stages of the model are as follows:
1. Voice Recognition [13]: The 'speech_recognition' library in used for this particular purpose. It uses Google Deep Learning [7] Speech Recognition function to process the voice input from the client device microphone. The snippet of the implementation code is as follows:

```
with speech_recognition.Microphone() as mic:
recognizer.adjust_for_ambient_noise (mic, duration=0.2)
    audio = recognizer.listen(mic)
    text = recognizer.recognize_google (audio)
    text = text.lower()
```

2. Tokenization and Word Filtering: The nltk library is used primarily for this part, which includes stop words, and POS Tags. The code snippet for importing the required library and functions is:

```
import nltk
from nltk.corpus import stopwords
from nltk.tokenize import word_tokenize
from nltk import pos_tag
```

There is a need for some preprocessing and pre-downloading a few elements for further processing. The code snippet for the same is as follows:

```
nltk.download('stopwords')
stop_words = set(stopwords.words('english'))
```

The code snippet for the tokenization and word filtering is as follows:

```
tokens = word_tokenize(text)
filtered_tokens = [word for word in tokens if word.casefold() not in stop_words]
posTags = pos_tag(filtered_tokens)
impWords = [item[0] for item in posTags if item[1] in impPostags]
```

3. Recommender algorithm: The algorithm needs the word vectors and the similarity function. The code snippet for importing, downloading, and pre-processing [13] the required from above is as follows:

```
import gensim.downloader as api
model = api.load('glove-wiki-gigaword-50')
word_vectors = model
from sklearn.metrics.pairwise import cosine_similarity
```

The data for the E-Commerce product categories is in JSON Object format. The structure of data for each product category in the json object is as follows:

```
"product": {
        "keyword_1": frequency_1,
        "keyword_2": frequency_2,
        .
        .
        .
        "keyword_n": frequency_n
}
```

One of the important things to keep in the equation while considering conversations is the actual motivations and the sentiment behind the statements. Hence, for basic sentiment analysis, the library "textblob" in python has been used. For the text that has been extracted, textblob sentiment analysis is applied to extract the polarity of the statements. It ranges from [-1, 1] where -1 implies complete negative emotions and +1 implies complete positive emotion. This factor is also used in the further recommendation process. The basic code used for the purpose is as follows:

```
from textblob import TextBlob
positivity = True
  if TextBlob(text).sentiment.polarity <0.2:
      positivity = False
```

It might be interesting to note that the parameter of distinguishing the positivity is kept at 0.2 rather than 0.0. This is done because a lot of peculiar cases emerge where slight negative motivation also obtains a polarity value greater than 0. For example, the statement "I don't want a dress" has a clear negative intent. However, the value returned is 0.136. Hence, considering such cases where clear positive intention is absent, the distinguishing value is set at 0.2.

There are two stages of comparison. The first one being with the headers and the second one being with the keywords. The code snippet for the first stage is as follows:

```
keys = list(data.keys())
  indices = dict()
  ind = 0
  for key in keys:
    indices[ind] = float(cosine_similarity([word_vectors[key.lower()]], [word_vectors[word]]))
    ind = ind + 1
  indices = dict(sorted(indices.items(), key=operator.itemgetter(1), reverse=False if positivity == False else True))
  indices = {key: indices[key] for key in list(indices)[:3]}
```

The comparison with the keywords also includes the additional factor of keyword frequency for the weighted average. The code snippet for the second stage is as follows:

```
vals = list(data.values())
  maxAvg=0
  maxInd = -1
  ind = 0
  indices = dict()
  for val in vals:
    avg = 0
    vals_length = 0
    for value in val.items():
      try:
        avg = avg + (value[1] * cosine_similarity([word_vectors[word]], [word_vectors[value[0]]])[0])
        vals_length = vals_length + value[1]
      except:
        continue
    avg = avg / vals_length
    indices[ind] = float(avg)
    if avg > maxAvg:
      maxInd = ind
      maxAvg = avg
    ind = ind + 1
  sorted_indices = dict(sorted(indices.items(), key=operator.itemgetter(1), reverse=False if positivity == False else True))
  indices = {key: sorted_indices[key] for key in list(sorted_indices)[:3]}
```

The program runs in the form of a web server, using flask. The product data is sent by the client to the server for comparison. After the processing, the top 3 relevant products, and the list of the filtered important words are sent back to the client, for further processing on the user's end.

**VI. RESULT**

The working of the conversational recommender model prototype for an E-Commerce marketing application can be displayed on a Flutter application and a React Website. These are the various stages of the working of the model:

Initially, the list of the product categories is displayed on top of the page, with a button which, on clicking, allows the user to input their conversational data.

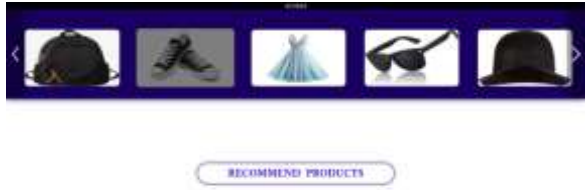

Figure 1: CVR Website Initial State

On clicking the button, the webapp actively reads the voice input. As an example, for processing the sentence "I need a new dress", these are the following products which are recommended.

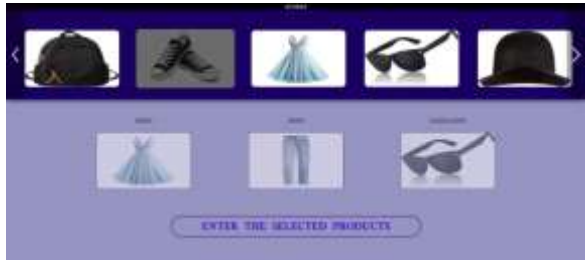

Figure 2: CVR Website after Recommendation

After the server returns the top three products, now the user's preference comes into the picture. Out of the three recommended products, the user can select which of the products he wanted to be recommended, in terms of personal preference.

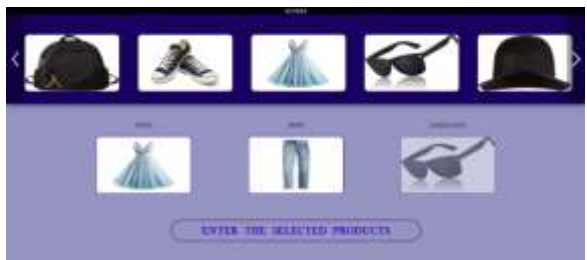

Figure 3: CVR Website after selection

For example's sake, only the first two categories are selected. After clicking the button on the bottom, two steps of further processing occur:

1) The frequency and the important words, returned by the server, are added to only those selected product categories, to increase their preference for the user. The updated product data is stored in local storage, so that it can be used whenever the user wants to make another recommendation.
2) The recommended products and the button at the bottom disappear, and the page is reverted to the initial status, ready for getting the voice input from the client device.

There might also be scenarios where the statement uttered by the user comes from a negative intent. Hence, taking another example where the input is "I don't want a new dress," these are the products that are recommended:

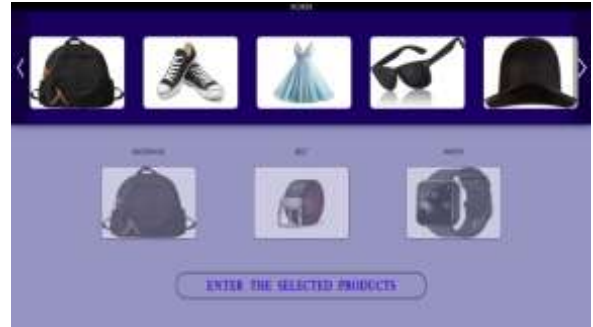

Figure 4: CVR Website after negative polarity input

## VII. FUTURE SCOPE

Voice recognition [13] and Natural Language Processing [2], using the technology of Deep Learning would prove to be highly effective in the development of a conversational recommender system. In terms of the model prototype, the recommender algorithm gives recommendation based on a limited range of product categories. In order to use the model for a real E-Commerce application, it would have to be adjusted in a few ways. One of them being the addition of brand names and various other specifications which might play a major factor in calculating the recommended products. Consequently, if the model is to be used for other purposes like movies, and music, the keywords would include specifications in that domain, which would also be taken into consideration.

All the product and keyword data are stored at the client side; hence it inherently is a safer method as compared to storing a large chunk of data all on the server side. Additionally, it helps in the personalization of recommendation. However, for the prototype, the data is stored in a relatively raw format without encryption applied. Hence, latest encryption methods would be applied to the personalized data as well to provide enhanced security measures.

Lastly, it is considered that as a prototype, the number of product categories and keywords will be limited. However, when considering enterprise level application dealing in terabytes of data, the sheer volume of categories and keywords would massively increase. Hence, to keep in mind the scalability factor, the number of keywords stored in each category would have to be capped, and only the most relevant keywords would have to be detected first (at an individual client-side level) and then retained accordingly, to optimize the storage and computation load.

## VIII. CONCLUSION

Conversational Recommender Systems represent a step forward in the evolution of recommendation systems. This is due to the analysis of words and meanings behind the conversations of users and understanding their subconscious behaviour with regards to the consumption of specific products; in the case of an E-Commerce application. Deep Learning techniques would prove to be appropriate for the purposes of Voice Recognition [13], Natural Language Processing [2], and Recommendation System modelling. The prototype of the conversational recommender system would demonstrate an improvement in the precision and relevance of recommendations over previous systems.

★ ★ ★